\documentclass[aps,showpacs,prd,preprintnumbers,nofootinbib,superscriptaddress]{revtex4-2}
\usepackage{graphicx}
\usepackage{url}
\usepackage{amsmath,amsthm,amssymb}
\usepackage{cancel}
\usepackage{slashed}

\usepackage[dvipsnames]{xcolor} % Must load first for [svgnames]

\definecolor{mpl-orange}{HTML}{FFA500}
\definecolor{mpl-green}{HTML}{008000}

\usepackage{afterpage}
\usepackage{amsfonts}
\usepackage{anyfontsize}
\usepackage{booktabs} % Required for better horizontal rules in tables
\usepackage{makecell}
\usepackage{mathtools}
\usepackage{multirow}
\usepackage{physics}
\usepackage[normalem]{ulem}
\usepackage{xspace} % Intelligent spacing for \institution and \department
\usepackage[english]{babel}
\usepackage[section]{placeins}

\usepackage{newtxtext,newtxmath}

\def\MadGraph{{\sc MG5\_aMC}\xspace}
\newcommand{\madgraph}{{\sc MG5\_aMC}\xspace}

\newcommand{\fastjet}{{\sc FastJet}\xspace}
\newcommand{\madanalysis}{{\sc Mad\-A\-na\-ly\-sis\,5}\xspace}
\newcommand{\pythia}{{\sc Pythia\,8}\xspace}

\newcommand{\hackanalysis}{{\sc HackAnalysis}\xspace}
\newcommand{\BSMArt}{{\sc BSMArt}\xspace}

\newcommand{\MET}{\ensuremath{E_{\text{T}}^{\text{miss}}}\ }
\newcolumntype{P}[1]{>{\centering\arraybackslash}p{#1}}
\DeclareMathOperator*{\argmin}{argmin}
\usepackage{makecell} \setcellgapes{2pt} %ANM: table row spacing
\usepackage[hidelinks]{hyperref}
\newcommand{\orcid}[1]{\begingroup
  \hypersetup{hidelinks}\href{https://orcid.org/#1}{\includegraphics[width=9pt]{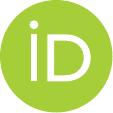}} \endgroup}

\pdfoutput=1

\begin{document}

\title{Improving smuon searches with Neural Networks}

\author{Alan S. Cornell\, \orcid{0000-0003-1896-4628}}
\email{acornell@uj.ac.za}
\affiliation{Department of Physics, University of Johannesburg, \vspace*{.05cm}\\
PO Box 524, Auckland Park 2006, South Africa.}
\author{Benjamin Fuks\,\orcid{0000-0002-0041-0566}}
\email{fuks@lpthe.jussieu.fr}
\affiliation{Laboratoire de Physique Th\'{e}orique et Hautes \'{E}nergies (LPTHE),\vspace*{.05cm}\\ UMR 7589, Sorbonne Universit\'{e} \& CNRS,\\ 4 place Jussieu, 75252 Paris Cedex 05, France}
\author{Mark D.  Goodsell\,\orcid{0000-0002-6000-9467}}
\email{goodsell@lpthe.jussieu.fr}
\affiliation{Laboratoire de Physique Th\'{e}orique et Hautes \'{E}nergies (LPTHE),\vspace*{.05cm}\\ UMR 7589, Sorbonne Universit\'{e} \& CNRS,\\ 4 place Jussieu, 75252 Paris Cedex 05, France}
\author{Anele M. Ncube\,\orcid{0000-0002-3312-5729}}
\email{ancube@uj.ac.za}
\affiliation{Department of Physics, University of Johannesburg, \vspace*{.05cm}\\
PO Box 524, Auckland Park 2006, South Africa.}

\begin{abstract}
We demonstrate that neural networks can be used to improve search strategies, over existing strategies, in LHC searches for light electroweak-charged scalars that decay to a muon and a heavy invisible fermion. We propose a new search involving a neural network discriminator as a final cut and show that different signal regions can be defined using networks trained on different subsets of signal samples (distinguishing low-mass and high-mass regions). We also present a workflow using publicly-available analysis tools, that can lead, from background and signal simulation, to network training, through to finding projections for limits using an analysis and {\tt ONNX} libraries to interface network and recasting tools. We provide an estimate of the sensitivity of our search from Run 2 LHC data, and projections for higher luminosities, showing a clear advantage over previous methods. 
\end{abstract}

\maketitle

\section{Introduction}

2012 saw the Standard Model (SM) vindicated with the confirmation of the Higgs boson by LHC’s ATLAS and CMS experiments, which was a testament to the effectiveness of particle physics models in predicting elementary particles’ properties and their interactions. With known SM processes as backgrounds, current experiments at the LHC are underway to search for signals of new physics processes predicted by a variety of Beyond the Standard Model (BSM) models fixing the shortcomings of the SM (see for instance Refs.~\cite{ATLAS} and \cite{CMS} for ATLAS and CMS searches for supersymmetry). The rigour of the ongoing searches depends not just on accurate Monte Carlo simulations, which provide phenomenological predictions, but also on the multivariate analysis used in background-signal classification, leading to the setting of unequivocal constraints on the masses of the predicted particles. Therefore, there exists frameworks such as \madanalysis \cite{Conte:2012fm,Conte:2018vmg} and \hackanalysis~\cite{Goodsell:2021iwc,Goodsell:2024aig} used for phenomenological purposes allowing to test signals from simulations and finding projections for exclusion limits (other frameworks include {\sc RIVET} \cite{Buckley:2010ar,Buckley:2019stt,Bierlich:2019rhm}, {\sc Contur} \cite{Butterworth:2016sqg,Buckley:2021neu}, {\sc ColliderBit} \cite{GAMBIT:2017qxg,GAMBIT:2018gjo,Kvellestad:2022uzz,GAMBIT:2023yih} and {\sc CheckMATE} \cite{Drees:2013wra,Dercks:2016npn,Desai:2021jsa}).

In the earlier days of the LHC operations, with a large jump in the available centre-of-mass energy between the Tevatron and Run 1, and then from Run 1 to Run 2, many BSM scenarios were immediately ruled out because they gave signal signatures associated with relatively very small background. However, now that no great leap in energy is available, we must assume that new physics is heavy or weakly-coupled enough that the signals are obscured by typically much larger backgrounds. In BSM searches where the background dominance is particularly stark, the sensitivity of an analysis algorithm to well-hidden signals is crucial. This has meant a much greater emphasis on developing techniques for increasing the discriminating power of an analysis. This can be done through finding new observables (such as in the pre-LHC era the derivation of the $m_{T2}$ observable \cite{Lester:1999tx,Barr:2003rg,Lester:2014yga}); searching for novel signatures (\textit{e.g.}\ long-lived particles); a careful analysis of existing observables; or, increasingly, the use of machine learning (ML) techniques to optimise background and signal separation (without necessarily requiring an underlying physical justification behind this separation). ML techniques are increasingly a versatile and powerful way to obtain the most information from the data available, with wide application in high-energy physics (see Ref.~\cite{hepmllivingreview} for an up-to-date bibliography), and a plethora of LHC analyses have already published results using them at every stage, see \textit{e.g.}\  Refs.~\cite{ATLAS:2019vwv,ATLAS:2020pcy,ATLAS:2020iwa,CMS:2020zge,Manganelli:2022whv}.

In the traditional cut-and-count approach, several signal regions are defined by a sequence of cuts, where events are rejected at each step based on specific criteria, and the number of events predicted at the end of each chain can be compared to data~\cite{annurev}. Some insight into the kinematics of the dominant background is needed to engineer cuts with the potential to effectively enhance the signal-to-noise ratio in a dataset that is subsequently passed through statistical analysis. For example, in searches for smuons within multi-lepton plus missing energy events, typical cut-and-count analyses (like the one in Ref. \cite{CMS:2018eqb}) specifically rely on a static jet veto forbidding the presence of jets with a transverse momentum $p^{j}_T$ greater than some threshold. Smuon production represents the quintessential low signal production rate scenario for which searches have remained inconclusive, motivating using it for tests and comparison of different search techniques. In this context, Ref.~\cite{Fuks:2019iaj} demonstrated potential improvements in sensitivity (across the hypothetical smuon-neutralino mass domain within the current capabilities of the LHC) by using dynamic jet vetoes~\cite{Pascoli:2018rsg, Pascoli:2018heg} instead of static ones. Initially, it was proposed to determine the cut threshold on the jet transverse momentum on an event-by-event basis, while more recently, event rejection was decided upon the relative global hadronic and leptonic activity in the events. This yielded increased discrimination power due to the differing relative amounts of hadrons and leptons in signal and background events. Subsequently, Ref.~\cite{Cornell:2021gut} explored the potential of a machine-learning approach using Boosted Decision Trees (BDTs) (see \textit {e.g.}\ Ref.~\cite{2024EPJST.tmp..292C} for a review) to the same model, although without publishing sensitivity projections. 

In this work, we extend the previous work on searching for {right-handed smuons, which exhibit the lowest cross-sections and have therefore been among the most challenging particles to search for. We propose a new search involving a neural-network (NN) discriminator as a final cut, and show that different signal regions can be defined using networks trained on different subsets of signal samples (distinguishing low-mass and high-mass regions). We also present a workflow using publicly-available analysis tools, that can lead from background and signal simulations to network training through finding projections for limits, and {\tt ONNX}\footnote{\url{https://onnx.ai/}} libraries to interface the network and recasting tools. In section \ref{SEC:DATASET} we describe the details of our search and the technology that we have developed for preparing the datasets required for training our NNs. In section \ref{SEC:TRAINING} we describe our proposed NN architecture, the training techniques, and the performance of the network. In section \ref{SEC:PROJECTIONS} we provide an estimate of the sensitivity of our search from Run 2 LHC data, and projections for higher luminosities, showing a clear advantage over both static and dynamic jet veto and BDT methods. We conclude in section~\ref{SEC:CONCLUSIONS}.

\section{Dataset preparation}
\label{SEC:DATASET}

\begin{table}[!t]
  \centering\renewcommand{\arraystretch}{1.3}
\begin{tabular}{c}
Preselection cuts\\ \hline
$\MET$  $> 100$ GeV \\
$\ge 2 $ isolated muons \\
Muons of opposite charge \\
$m_{\mu \mu} > 20$ GeV and $|m_{\mu\mu} - m_Z| > 15 $ GeV\\[.4cm] 
Additional cut imposed on the dataset \\ \hline
$m_{T2} > 90$ GeV \\[.4cm]
Cuts for comparison with Ref.~\cite{Fuks:2019iaj}, not imposed on dataset \\ \hline
$p_{T} (\mu_1) > 50\ \mathrm{GeV}, p_{T} (\mu_2) > 20\ \mathrm{GeV}$ \\
\end{tabular}
\caption{\label{TAB:preselection} Selection cuts imposed in our analyses. Only events passing the preselection cuts are written to disk. The $m_{T2}$ cut is imposed directly on the dataset prior to training the network; the final lepton $p_T$ cut is not imposed in our analysis, but only shown in our cutflows for comparison with other work.}
\end{table}

In order to train NNs to achieve better performance than standard cut-and-count searches, or even ML techniques involving BDTs, we must first generate an appropriate dataset. In order to do this we created a ``pseudo-analysis'' in \madanalysis~\cite{Conte:2012fm,Conte:2018vmg} and \hackanalysis \cite{Goodsell:2021iwc,Goodsell:2024aig} to apply a series of basic cuts and then collate features of the remaining events into compressed CSV data files. The data stored will be described in the next section. The main backgrounds for smuon pair production and decay into a muon pair and missing energy originate from $\ell\ell\ell\nu$ and $\ell\ell\nu\nu$ events (chiefly through $WW$ and $WZ$ production) and $t\overline{t}$ events. We therefore simulate high-statistics samples of these events with \MadGraph~\cite{Alwall:2014hca}, in addition to a selection of signal samples with different right-handed smuon and neutralino masses with the decay $\tilde{\mu}_R \rightarrow \tilde{\chi}_1^0 + \mu$ enforced at $100\%$, using the (conventional) UFO~\cite{Darme:2023jdn} implementation of the Minimal Supersymmetric Standard Model developed in Ref.~\cite{Duhr:2011se}. Event simulation is performed by convoluting leading-order (LO) matrix elements at a centre-of-mass energy of 13~TeV with the LO set of NNPDF3.1 parton densities~\cite{NNPDF:2014otw, Buckley:2014ana}, that we then match with parton showering and hadronisation as modelled by \pythia~\cite{Bierlich:2022pfr}. Event reconstruction is performed with \fastjet~\cite{Cacciari:2011ma} and the anti-$k_T$ algorithm~\cite{Cacciari:2008gp}. The detector simulation function in \hackanalysis is called {\tt FATJET} which uses the default (perfect) detector but with large radius jets. 

Employing ML poses an interesting challenge compared to traditional analyses, because we must not only collect enough signal events to have a statistically accurate measurement of the efficiency of our final cut, but we need sufficient events that should pass all cuts to train the NN: this means that we need far more data samples than would typically be required. Moreover, in order to avoid overtraining on a single data point, it is necessary to select several signal samples. At the same time, it is necessary to have a balanced dataset, so we require a large number of background samples too. As a result, we restrict ourselves to leading-order simulations of both backgrounds and signals (for a dedicated analysis with access to the full detector simulation, then higher-order simulations of the background events would be standard); this has the added advantage that we are not biasing the NN training on some features of the differences between higher-order and leading-order simulations between background and signals.

Our pseudo-analysis broadly follows the pre-selection cuts of Ref.~\cite{Fuks:2019iaj} which we list in table \ref{TAB:preselection}, except that we specifically target muons rather than any light lepton. We define signal electrons with $p_T > 7$ GeV and $|\eta| < 2.47,$ muons with $p_T > 7$ GeV and $|\eta| < 2.5,$ taus with $p_T > 20$ GeV and $|\eta| < 2.5,$  and photons with $p_T > 10$ GeV and $|\eta| < 2.37.$ All are required to fulfill isolation requirements such that the sum of energy in a cone of $\Delta R =0.4$ around them is less than $20\% $ of the object $p_T.$ Signal jets are clustered with a wide radius parameter $R = 1$ (which helps to capture jets from top quark decays), and required to have $p_T > 15$ GeV  and $|\eta| < 2.8.$ We remove jets in a cone of $\Delta R =0.4$ around muons and photons, and $\Delta R = 0.2$ around electrons and taus; then we remove electrons within $\Delta R = 0.4$ of jets, and photons within $\Delta R =0.4$ of electrons and muons. 

\begin{table}[t!]\centering
\makegapedcells
\begin{tabular}{P{0.1\linewidth}|P{0.15\linewidth}P{0\linewidth}P{0.1\linewidth}|P{0.15\linewidth}P{0\linewidth}P{0.15\linewidth}|P{0.15\linewidth}}
\multicolumn{2}{c}{low-level}                                    & & \multicolumn{2}{c}{high-level}                                     & & \multicolumn{2}{c}{derived}\\
Variable          &  Group                                      & & Variable                                & Group                   & & Variable                                 & Group \\ \cline{1-2}\cline{4-5}\cline{7-8}
$n_{\ell}$        & \multirow{2}{8em}{\centering final state particle count} & & $m_{\mu\mu}$                  & \multirow{3}*{\centering masses} & & $S_T$ & {\centering} sum of lepton $p_T$ \\ \cline{7-8}
$n_{j}$           &                                                & & $m_{CT}$                                &                            & & $\vert\Delta\eta_{\mu_1, \mu_2}\vert$ & \multirow{2}{8em}{\centering muon angular separation} \\\cline{1-2}
%$\vert\Delta\phi_{\mu_1, \mu_2}\vert$  & \\ \cline{1-2}
$p^{\ell_1}_T$    & \multirow{2}{8em}{\centering lepton momenta}   & & $m_{T2}$                                &                            & &                                          $\vert\Delta\phi_{\mu_1, \mu_2}\vert$ & \\ \cline{4-5}\cline{7-8}
$p^{\ell_2}_T$    &                                                & & $\slashed{E}_T$                         & \multirow{2}{8em}{\centering missing transverse energy} & & $p^{\mu_1}_T/p^{\mu_2}_T$ & \multirow{8}{8em}{\centering ratio of leptonic and hadronic observables}\\ \cline{1-2}
$\eta^{\ell_1}$   & \multirow{4}{8em}{\centering lepton angles}    & &                                         &                            & & $p^{\mu_1}_T/p^{j_1}_T$                 &  \\ \cline{4-5}
$\eta^{\ell_2}$   &                                                & & $\phi(p_T^{\rm miss})$                                 & \multirow{2}{8em}{\centering missing traverse momentum azimuthal angle}  & &  $p^{\mu_1}_T/H_T$ & \\
$\phi^{\ell_1}$   &                                                & &                                  &                            & & $S_T/H_T$                                & \\
$\phi^{\ell_2}$   &                                                & &                                         &                            & & $p^{\mu_2}_T/p^{j_1}_T$                 & \\ \cline{1-2}
$p^{j_1}_T$       & \multirow{2}{8em}{\centering jet momenta}      & &                                         &                            & & $S_T/p^{j_1}_T$                          & \\ \cline{4-5}
$p^{j_2}_T$       &                                                & & $H_{T}$                  & \multirow{1}{8em}{\centering sum of hadronic observables} & & $p^{\mu_1}_T/p^{j_1}_T$ & \\ \cline{1-2}
$\eta^{j_1}$      & \multirow{4}{8em}{\centering jet angles}       & &                                         &                            & &                                          & \\ 
$\eta^{j_2}$      &                                                & &                                         &                            & &                                          & \\
$\phi^{j_1}$      &                                                & &                                         &                            & &                                          & \\
$\phi^{j_2}$      &                                                & &                                         &                            & &                                          & \\
\end{tabular}
\caption{\label{TAB:features} Kinematic variables used as input features to the NNs. Here $\ell$ represents electrons and muons with separate entries for each (so there are two variables for every entry labelled with $\ell$). Hence there are 21 variables in the left panel, 6 in the middle, and 10 in the right panel. }
\end{table}

For events passing the preselection cuts, the low-level and high-level features in the first and second panels of table~\ref{TAB:features} are assembled into a vector, which is written into a comma-separated values file, directly in compressed format. Separate files are created for each process, and the files from different runs combined afterwards. The set of variables includes lepton and jet multiplicities $n_\ell$ (actually given separately as $n_e, n_\mu$ being the number of electrons and muons respectively) and $n_j$, the transverse momenta, pseudo-rapidities and azimuthal angles of the two leading jets, two leading electrons (if any)\footnote{Null entries are represented by a large negative number ($-10000$).} and two leading muons (if any), the invariant mass of the lepton pair $m_{\mu\mu}$, the contratransverse mass $m_{CT}$ of the system~\cite{Tovey:2008ui}, the $m_{T2}$ variable, the missing transverse energy $\slashed{E}_T$, the hadronic activity $H_T$ defined as the scalar sum of the jet $p_T$, and the azimuthal direction of the missing transverse momentum. During the processing of these files, `derived features' (last panel of the table) are calculated from the low-level and high-level ones and fed to the network, as described in the next section.

\begin{table}\renewcommand{\arraystretch}{1.2}\setlength\tabcolsep{12pt}
\begin{tabular}{c|ccc}
Cut & $t\overline{t}$ & $\ell \ell \nu\nu$ & $\ell\ell\ell\nu$  \\ \hline
Initial & 1320500.0 & 116760.0 & 22240.0\\
Basic Cuts & 318087.1 & 11137.5 & 1778.4\\
$\ge 2$ Muons & 142824.5 & 6940.8 & 1672.1\\
Opposite Charge & 142562.7 & 6940.3 & 1639.7\\
$m_{\mu\mu}$& 106399.8 & 3180.1 & 450.9\\ \hline
$m_{T2}$ & 376.9 & 307.4 & 57.4\\
Lepton $p_T$ & 174.6 & 188.1 & 44.0\\ 
\end{tabular}
\caption{\label{TAB:BackgroundYields} Background yields for the different processes considered, for $139~\mathrm{fb}^{-1}$ at $13$ TeV.} 
\end{table}

We simulated three datasets of background events. The first consists of $t \overline{t}$ production where both tops decay muonically (that is, to a muon, neutrino and $b$-quark) (with the decay performed within \madgraph to preserve angular correlations); the second consists of the process $pp\to \mu^+\mu^-\nu\overline{\nu}$, which we denote $\ell\ell\nu\nu$; and the third is related to the process $pp\to \mu^+ \mu^- \ell^\pm\nu_\ell/\overline{\nu}_\ell$, which we denote $\ell \ell \ell \nu$. Each of these processes is simulated with up to two additional jets, and merged using the MLM scheme~\cite{Mangano:2006rw, Alwall:2008qv} with a matching scale of $60$ GeV. We take the cross-section for the $p p \rightarrow t\overline{t} \rightarrow \mu \overline{\mu} b \overline{b} \nu\overline{\nu}$ process to be $9.5$~pb, which corresponds to the measured values from ATLAS/CMS of around $830$ pb multiplied by a branching ratio of $(0.1071)^2$, in agreement with the most precise available theory predictions~\cite{Czakon:2013goa}. For the $\ell \ell \nu \nu$ process we have $840$~fb, and for the $\ell \ell \ell \nu$ process we have $160$~fb, which both correspond to the values reported in \madgraph multiplied by a constant $K$-factor of $1.2$. We report the yields obtained for a luminosity of $139 \mathrm{fb}^{-1}$ after each cut in table~\ref{TAB:BackgroundYields}. We estimated the systematic uncertainties on these values using the systematics feature of \hackanalysis to be of $10\%$ on the final cut.

We know from Ref.~\cite{Fuks:2019iaj} that a jet veto will reduce the $t\overline{t}$ background by a factor of around 100, which would leave us with only of order one event; this is not surprising, since we are using fat jets which are typically used to tag top quarks. On the other hand, the reduction for the $\ell\ell\nu\nu$ process is only of order $50\%$, which still leaves us with a large number of background events; and of course there would be a significant reduction in our signal. Since our aim is to improve over this approach via a NN, we require a large enough sample after the $m_{T2}$ cut. Therefore we simulated of order $500$M $t\overline{t}$ events, yielding $160$M events after merging, and of order $44$k after the $m_{T2}$ cut. For the $\ell\ell\nu\nu$ case (potentially our most important/most irreducible background) we obtained  $60$M events after merging, and $160$k after the $m_{T2}$ cut. Finally, for the $\ell\ell\ell\nu$ case we obtained $8$M events after merging, and $22$k after the $m_{T2}$ cut. To simulate these large numbers of events, we used \hackanalysis running on multiple nodes, each using $8$ cores, and driven by \BSMArt \cite{Goodsell:2023iac,Faraggi:2023jzm,Goodsell:2024aig}. 

\begin{table}\renewcommand{\arraystretch}{1.6}\setlength\tabcolsep{12pt}\centering
\begin{tabular}{c|c|c|c}
Cut & $(m_{\tilde{\mu}},m_{\tilde{\chi}_1^0}) = (200,150)$& $(m_{\tilde{\mu}},m_{\tilde{\chi}_1^0}) = (600,0)$& $(m_{\tilde{\mu}},m_{\tilde{\chi}_1^0}) = (500,300)$ \\ \hline
Initial Events & $1148.9$ ${}^{+ 6.0\%}_{-5.0\%}$ (syst) & $11.6$ ${}^{+ 14.7\%}_{-8.1\%}$ (syst) & $27.6$ ${}^{+ 12.8\%}_{-7.5\%}$ (syst)\\
Basic Cuts & $289.4$ ${}^{+ 5.8\%}_{-5.3\%}$ (syst) & $10.9$ ${}^{+ 0.0\%}_{-0.1\%}$ (syst) & $23.3$ ${}^{+ 0.2\%}_{-0.2\%}$ (syst)\\
$\ge 2$ Muons & $100.1$ ${}^{+ 7.6\%}_{-6.5\%}$ (syst) & $6.9$ ${}^{+ 0.5\%}_{-0.4\%}$ (syst) & $14.0$ ${}^{+ 0.4\%}_{-0.5\%}$ (syst)\\
Opposite Charge & $100.1$ ${}^{+ 7.6\%}_{-6.5\%}$ (syst) & $6.9$ ${}^{+ 0.5\%}_{-0.4\%}$ (syst) & $14.0$ ${}^{+ 0.4\%}_{-0.5\%}$ (syst)\\
$m_{\mu\mu}$ & $73.3$ ${}^{+ 7.7\%}_{-6.5\%}$ (syst) & $6.8$ ${}^{+ 0.5\%}_{-0.4\%}$ (syst) & $13.3$ ${}^{+ 0.4\%}_{-0.5\%}$ (syst)\\ \hline
$m_{T2}$ & $4.8$ ${}^{+ 8.0\%}_{-6.9\%}$ (syst) & $6.0$ ${}^{+ 0.8\%}_{-0.5\%}$ (syst) & $10.7$ ${}^{+ 0.9\%}_{-0.6\%}$ (syst)\\ 
Lepton $p_T$ & $1.8$ ${}^{+ 6.2\%}_{-5.4\%}$ (syst) & $5.9$ ${}^{+ 0.8\%}_{-0.5\%}$ (syst) & $10.2$ ${}^{+ 1.0\%}_{-0.7\%}$ (syst)\\
\end{tabular}
\caption{\label{TAB:SignalCutflows} Cutflows for representative signal points, showing the predicted number of events after each cut for a luminosity of $139~\mathrm{fb}^{-1}$, with the systematic errors corresponding to scale variations.}
\end{table}

Signal samples were obtained in the same manner for a selection of training points. Since the signal typically survives the cuts with a high efficiency, it was only necessary to simulate $240$k events per parameter point (leading to roughly $100$k after merging). We show three illustrative cutflows in table~\ref{TAB:SignalCutflows}. There will clearly be an upper bound on the sensitivity to be expected, which corresponds to a configuration where about one event passes our preselection cuts. This occurs in two situations, first for large smuon masses and second for highly compressed spectra (where the smuon and neutralino are close in mass). For instance, for a $600$~GeV smuon, we get a cross-section around $0.08~\mathrm{fb}$. This leads to approximately $12$ events before any cut, which reduces to $6$ after the preselection in the case of massless neutralinos; this provides the upper bound in sensitivity that will be used to define our target mass range, which is illustrated in figure~\ref{FIG:Xs}. 

\begin{figure}
\includegraphics[width=0.45\textwidth]{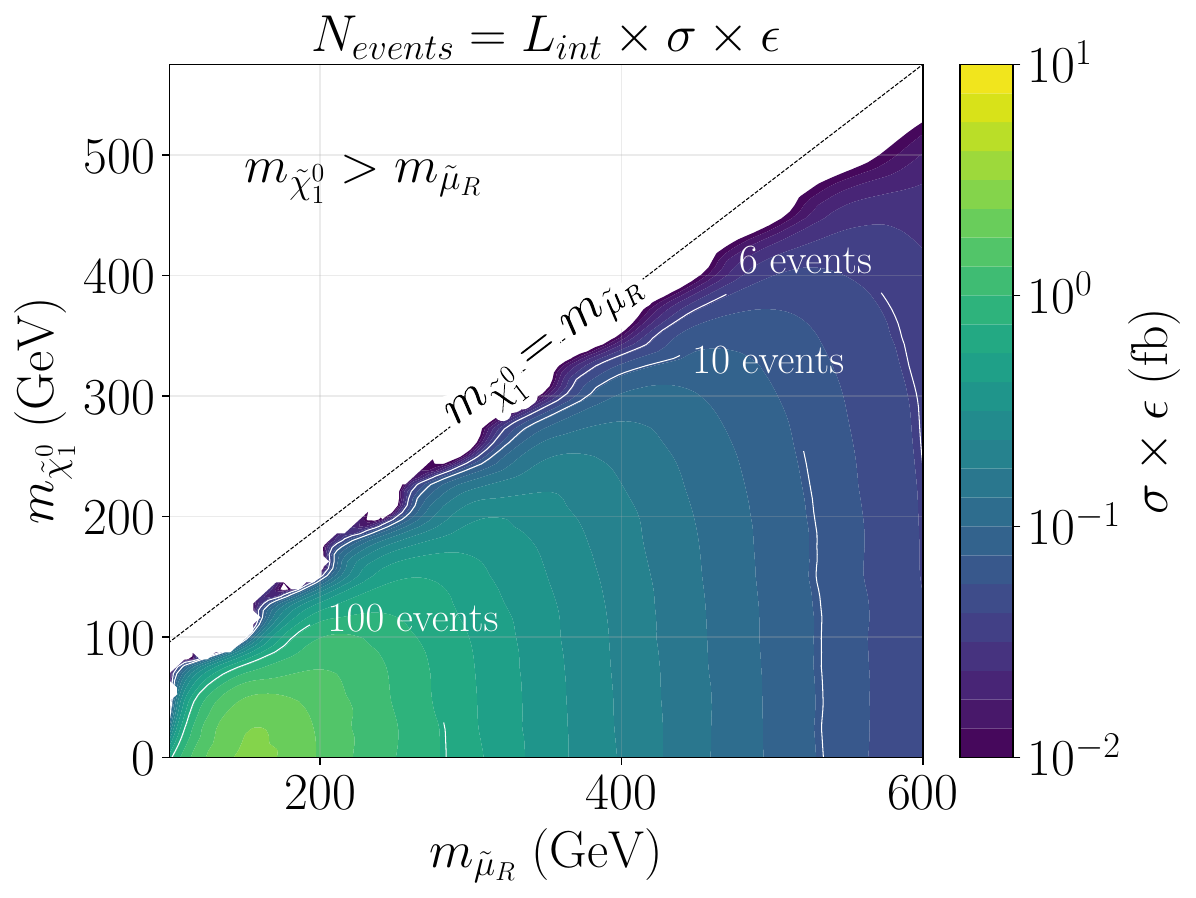}
\caption{\label{FIG:Xs}Fiducial signal cross-sections (and the corresponding number of events for a luminosity of 139~fb$^{-1}$) as a function of the smuon and neutralino mass, and after the preselection cuts. The mass range of interest is defined by the 6-event isoline.}
\end{figure}

\section{Network training}
\label{SEC:TRAINING}

\subsection{Basics of neural networks}

The task involved in discriminating the data due to BSM processes (in this case the production of smuons) from the data due to SM processes, whose production rate generally exceeds that of the BSM processes, is an imbalanced classification problem that can be solved using NN-based classifiers. The solution that the NN provides is a highly discriminative feature $f(x_{\alpha}) \in \mathbb{R}$, a transformation of the event observables $x_{\alpha} \in \mathbb{R}^d$ that most effectively separates events associated with signals from events associated with known backgrounds in the $d$-dimensional space of $x_{\alpha}$ \cite{ChathamStrong_2020}. Specifically, the event observables are the features listed in table \ref{TAB:features}. The optimal $f(x_{\alpha})$ that is learned from the training data generated from Monte-Carlo event generation (as described in section \ref{SEC:DATASET}) is given generally by:
\begin{equation}
\label{NN equation}
    f(x_{\alpha}) = \sigma^L_{\delta}\left[\sum_{\gamma} w^{L-1}_{\delta\gamma}\sigma^{L-1}_{\gamma}\left(\ldots \sum_{\beta} w^2_{\gamma\beta}\sigma^{1}_{\beta}\left(\sum_{\alpha} w^1_{\beta\alpha} x_{\alpha} + b^1_{\beta}\right) + b^2_{\gamma} \ldots\right) + b^{L-1}_{\delta}\right].
\end{equation}
Here the superscripts $\ell = 1, 2, \ldots L-1, L$ index layers of the NN. For any given layer, the quantities $w^{\ell}_{\beta\alpha}$ (and similarly the quantities $w^{\ell}_{\gamma\beta}$, $w^{\ell}_{\delta\gamma}$, ...) represent tunable weights and $b^{\ell}_{\beta}$ (or $b^{\ell}_{\gamma}$, $b^{\ell}_{\delta}$, ...) denote adaptable biases. Additionally, $\sigma^{\ell}_{\beta}$ (and also $\sigma^{\ell}_{\gamma}$, $\sigma^{\ell}_{\delta}$, ...) denote activation functions that are the last operations in a given NN layer. 

The loss $\mathcal{L}$ estimates the overall deviation of the NN from an optimal state. In binary classification, binary cross-entropy is the standard choice of loss function. Minimisation of the loss function results in the optimisation of the NN's predictions according to $f(\theta^*): \theta^* = \argmin_{\theta}$ $\mathcal{L}(\theta)$, where $\theta$ collectively denotes the adaptable weights and biases of the NN as $\theta = \{w^{\ell}_{\alpha\beta}, b^{\ell}_{\beta}\}^{L}_{\ell = 1}$. The optimisation is executed by an algorithm that uses derivatives $\nabla_{\theta} \mathcal{L}$ to update $\theta$ in the direction toward the minimum of $\mathcal{L}$. Standard optimisation algorithms include Stochastic Gradient Descent (SGD) with momentum method, which we used to tune our NNs.

\subsection{Preprocessing}
To implement NNs and carry out data wrangling, we have used {\tt pytorch} and other existing standard libraries for data preprocessing \cite{10.5555/3454287.3455008, scikit-learn, mckinney-proc-scipy-2010, harris2020array}. Having acquired data from Monte-Carlo event generation and implemented preselection cuts as detailed in section II, the simulated data was preprocessed for NN training through the following steps:
\begin{itemize}
    \item We determine 10 additional derived features, including in particular the sum and ratio of leptonic and hadronic observables, as given in the right panel of table~\ref{TAB:features}, that supplement the 27 pre-existing low- and high-level features.
    \item The signal data is split into two sets, namely low and high smuon mass events. These sets refer to scenarios with $m_{\tilde\mu_R} < 450$~GeV and $m_{\tilde\mu_R} > 450$~GeV, respectively. We train two NNs, one for each of the smuon mass regions.  
    \item We rescale the event weights by first multiplying by a factor such that the mean weight per event equals the cross-section times luminosity, and then dividing by the size of the dataset. This means that we can compute the expected number of events passing our NN cut by just summing their (rescaled) weights.
    \item A subset of the simulated signal samples (\textit{i.e}.\ 18 out of 154 benchmark points selected in the parameter space) are combined with the simulated background samples, and then were split into a training, validation and test sets according to the ratio 8:1:1. The training data was balanced to get equal numbers of events for each signal and background component. Standardisation was subsequently applied to all three sets using the mean and standard deviation of the balanced training set.
\end{itemize}

\subsection{Neural Network architecture}
\begin{table}[t!]\centering\setlength\tabcolsep{6pt}
\makegapedcells
\begin{tabular}{p{0.2\linewidth}|P{0.45\linewidth}|P{0.2\linewidth}}
Hyperparameter                                                                & Description & Selection  \\ \hline
\vfill Nodes ($n_{1}$-$n_{2}$-\ldots-$n_{L}$)\vfill                           & \vfill Number of elements $n_{\ell}$ of each NN layer $\ell = 1, 2, \ldots, L$\vfill & \vfill 30-50-50-1\vfill\\
\vfill Weight initialisation\vfill                                            & \vfill Initialisation of the tunable NN parameters\vfill & \vfill $\mathcal{N}(\mu= 1,\sigma^2 = 0.04)$\vfill\\
Activation function                                              & A linear or non-linear transformation of nodes of the NN &  ReLU\\
Optimisation algorithm                                                        & Drives NN learning & SGD\\
Learning rate ($lr$), weight decay ($\lambda$), momentum ($\mu$)              & \vfill Parameters of the SGD algorithm\vfill & \vfill $lr = 0.1, \lambda = 10^{-4}, \mu = 0.1$\vfill \\ 
Batch size  & Number of training points used in each weight update step & 32\\
Epochs                                                                        & The number of NN training rounds & 500\\ 
\end{tabular}
\caption{\label{TAB:Neural Network Architecture} Hyperparameters used in our {\tt pytorch} NN model.}
\end{table}

We employ a simple feed-forward three-layer NN architecture consisting of 37 input nodes (corresponding to the 27 low/high-level and 10 derived features\footnote{In fact we actually used 38 input nodes, with one input always set to zero corresponding to a feature that we deleted.}), connected to two sequential hidden layers of 50 nodes and one output layer of a single node. We use a sigmoid activation function to connect layers, as expressed by Eqn. \eqref{NN equation}. Lastly, the NN parameters are initialised by setting the components of $w^\ell_{\alpha\beta}$  and $b^{\ell}_{\alpha}$ to values sampled from a normal distribution $\mathcal{N}(\mu, \sigma^2)$ with mean $\mu = 1$ and standard deviation $\sigma = 0.2$. See table~\ref{TAB:Neural Network Architecture} for further details.

\subsection{Training}

\begin{table}[t!]\centering\setlength\tabcolsep{8pt}\renewcommand{\arraystretch}{1.25}
\makegapedcells
\begin{tabular}{P{0.1\linewidth}|P{0.2\linewidth}|P{0.3\linewidth}|P{0.2\linewidth}}
Threshold  & Threshold-induced rates & Model tendency & Expected predictions \\ \hline
\multirow{2}{5em}{0.5}        & high FPR  &  models trained to minimise FPR (``precision-oriented'') & lower FPR than in ``recall-oriented'' models\\
                              & high TPR  &  model trained to maximise TPR, but less effectively than for a threshold of 0.7  & lower TPR than in ``recall-oriented'' models\\ \hline
\multirow{2}{5em}{0.7}        & high FNR  &  model trained to minimise FNR (``recall-oriented'')     & higher FPR than in ``precision-oriented'' models\\
                              & low TPR   &  model trained to maximise TPR more effectively than for a threshold of 0.5  & higher TPR than in ``precision-oriented'' models\\
\end{tabular}
\caption{\label{TAB:Thresholds} Effect of threshold-setting during training on the TPR and FPR of the final data analysis.}
\end{table}
After data pre-processing, our NNs were trained over 500 epochs using the SGD algorithm with momentum. We refer to table~\ref{TAB:Neural Network Architecture} for our set-up of the SGD optimiser. An average score of several metrics relevant to binary classification ($f_{\beta}$-score, ROC AUC, precision and AUPRC), that was determined during training based on the validation set, was then used to determine an approximate best model. Note that for these metrics, which require setting a classification threshold, choosing a lower threshold in training (here 0.5) than in analysis (here 0.7) resulted in fewer true positives (TP) and false positives (FP) compared to choosing an equivalently high threshold in training and in analysis (0.7 in both cases). To explain this, the trade-off between precision and recall is a potential cause; precision and recall are defined as:
\begin{equation}
    \text{Precision} = \frac{\text{TP}}{\text{TP} + \text{FP}},\quad \text{Recall} = \frac{\text{TP}}{\text{TP} + \text{FN}}. 
\end{equation}
In other terms, recall (precision) quantifies the fraction of relevant (retrieved) items that are retrieved (relevant), the symbol FN representing the false negatives. In table~\ref{TAB:Thresholds}, we deduce from the ``priorities'' set on precision and recall, namely the constraints imposed on the true positive rate (TPR), false positive rate (FPR), true negative rate (TNR) and false negative rate (FNR), the expected behaviours of what we call ``precision-oriented'' and ``recall-oriented''. As such the ``precision-oriented'' models that we train from setting the threshold to 0.5 during training are optimised to reduce backgrounds (FPs) at the cost of a lower signal acceptance (TPs). Conversely, the ``recall-oriented'' models obtained from setting the threshold to 0.7 both during training and analysis would give sub-optimal background rejection and better signal acceptance. In analyses, our ``precision-oriented'' NN models are the preferred trade-off to prioritise reducing the backgrounds. 

\begin{figure}
\includegraphics[width=0.8\textwidth]{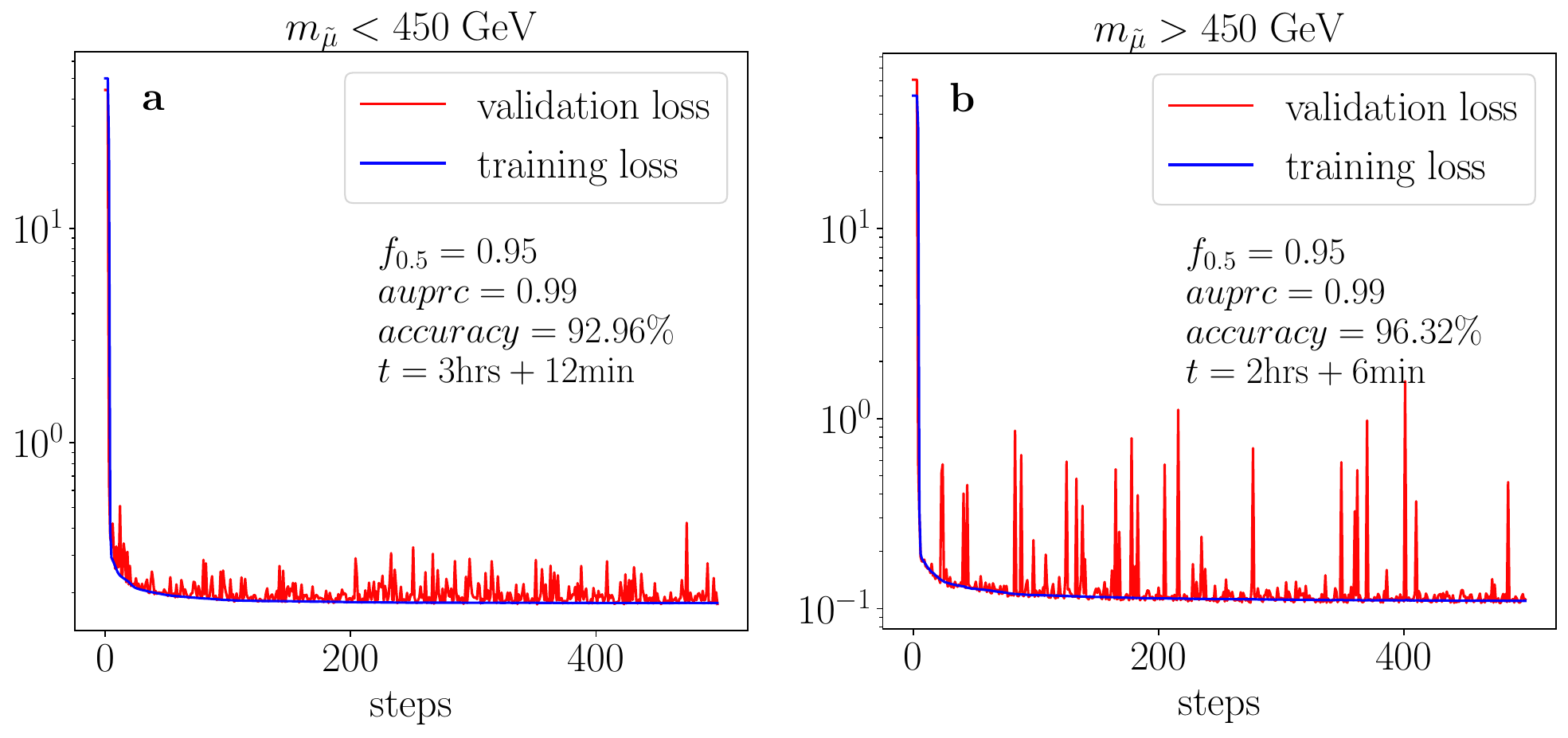}
\caption{\label{FIG:losses} The low-mass (signal samples such that $m_{\tilde{\mu}} < 450 $~GeV) and high-mass ($m_{\tilde{\mu}} > 450$~GeV) network loss histories.}
\end{figure}

The loss histories generated while training the low-mass and high-mass NNs are shown in figure~\ref{FIG:losses}. The spikes seen in the validation loss are due to the fewer data points in the validation set than in the training set, which is more pronounced for the high-mass network that uses a smaller dataset than the low-mass one. Regardless, the similarity in the training and validation losses throughout training indicates that the models generalise well, which is corroborated by the high scores on the metrics evaluated using the test set. Training the NNs took approximately three and two hours, respectively, for the low-mass and high-mass networks.

Some results pertaining to NN performance that are worth noting were obtained before we used the approach of training two separate networks dedicated respectively to a low-mass and high-mass signal. In this earlier try, all the backgrounds and signal samples were combined and trained on a single network. These results gave us some insights into the NN's ability to automatically learn information from low-level features that are provided by the high-level and derived features. In this case, three different NN setups were tested: a first NN was trained on only low-level features, while a second network was trained only on high-level and derived features. Finally, a third network was trained on an aggregate of all features. The performances measured during training of each of these three NN setups in terms of accuracy (ACC) and $f_1$ score were of $\mathrm{ACC}=94\%$ $f_1=0.96$ for the NN using only low-level features, $\mathrm{ACC}=96\%$ and $f_1=0.97$ for the NN using only high-level and derived features, and $\mathrm{ACC}=96\%$ and $f_1=0.97$ for the NN using all features. From these values, we see that there is $2\%$ improvement in accuracy due to using the high-level features exclusively or in combination with low-level features. 

\begin{table}[t!]\centering\setlength\tabcolsep{8pt}\renewcommand{\arraystretch}{1.25}
\begin{tabular}{P{0.1\linewidth}P{0.1\linewidth}|P{0.1\linewidth}P{0.3\linewidth}P{0.1\linewidth}}
\multirow{2}{*}{$m_{\tilde{\mu}_{_R}}$ (GeV)} & \multirow{2}{*}{$m_{\tilde{\chi}^0_1}$ (GeV)} & \multicolumn{3}{c}{ROC AUC} \\
 &                      & low & high + derived & all  \\[0.5em]
\hline
200                    & 0                      & 0.972 & 0.988        & 0.988\\ 
300                    & 0                      & 0.991 & 0.996        & 0.997\\
500                    & 0                      & 0.998 & 0.999        & 0.999\\
\end{tabular}\caption{\label{TAB:Performances} ROC AUCs of 3 different NN discriminators trained on low-level features, (``low''), high-level and derived features (``high+derived'') or an aggregate of all of them (``all'').} 
\end{table}

On the other hand, we measured the test performance (after training) using the ROC AUC metric, representative results for a few benchmark scenarios with a massless neutralino being reported in table~\ref{TAB:Performances}. The relative test performances of the three different NN setups concur with the accuracies, the ROC AUCs being greater for the NNs employing high-level inputs. This outcome is expected given that high-level and derived features have, by design, greater discriminative power than low-level features. However, as indicated in Ref.~\cite{pmlr-v42-cowa14}, optimising the ROC AUC is not equivalent to optimising the overall sensitivity of the NN. One reason for this is that while the ROC AUCs gives us an estimate of the general performance of the classifier, it can be biased towards the larger size background class. Overall, the difference in performance is not drastic between NNs using only low-level and only high-level plus derived features or all features at the same time, which indicates that all NNs, and in particular the ones using only low-level features, automatically learn the insights provided by the high-level features. A similar conclusion is reached in Ref.~\cite{Baldi:2014kfa} that relies also on NNs and, more pertinently, the work of Ref.~\cite{Cornell:2021gut} that relies on BDTs to address the same physics case.

\subsection{Performance}
\begin{table}[t!]\centering\setlength\tabcolsep{10pt}\renewcommand{\arraystretch}{1.3}
\begin{tabular}{P{0.1\linewidth}P{0.1\linewidth}|P{0.05\linewidth}P{0.1\linewidth}|P{0.05\linewidth}P{0.1\linewidth}|P{0.05\linewidth}P{0.1\linewidth}}
\multirow{2}{*}{$m_{\tilde{\mu}_{_R}}$ (GeV)} & \multirow{2}{*}{$m_{\tilde{\chi}^0_1}$ (GeV)}  & \multicolumn{2}{c|}{NN} & \multicolumn{2}{c|}{BDT} & \multicolumn{2}{c}{Static jet veto}\\
& & $s$ & $\varepsilon$ & $s$ & $\varepsilon$ & $s$ & $\varepsilon$\\
\hline
\multirow{1}{0em}{100} &  0                     &  19.5 & 21.7\%       &  53.3 & 59.5\%        &  37.2 & 41.6\% \\
                       &  25                    &   3.5 & 16.7\%       &  11.9 & 56.3\%        &   8.0 & 37.8\% \\[0.2em]
\multirow{1}{0em}{200} &  0                     & 172.1 & 72.3\%       & 214.5 & 90.1\%        & 129.6 & 54.4\% \\
                       &  50                    & 146.4 & 68.2\%       & 189.5 & 88.3\%        & 113.6 & 52.9\% \\
                       &  100                   &  71.0 & 51.6\%       & 109.9 & 79.9\%        &  67.4 & 49.0\% \\
                       &  150                   &   0.4 & 9.2\%       &   2.5 & 51.7\%        &   0.5 & 11.2\% \\[0.2em]
\multirow{1}{0em}{300} &  0                     &  74.3 & 87.6\%       &  82.1 & 96.8\%        &  45.2 & 53.3\%) \\
                       &  50                    &  70.5 & 86.7\%       &  78.4 & 96.4\%        &  43.1 & 53.0\% \\
                       &  100                   &  64.8 & 84.0\%       &  73.3 & 95.1\%        &  40.4 & 52.4\% \\
                       &  200                   &  23.7 & 58.2\%       &  33.7 & 83.0\%        &  19.3 & 47.5\% \\
                       &  250                   &   0.5 & 13.5\%       &   1.9 & 55.5\%        &   0.5 & 15.3\% \\[0.2em]
\multirow{1}{0em}{400} &  0                     &  29.3 & 91.4\%       &  31.6 & 98.5\%        &  16.6 & 51.7\% \\
                       &  200                   &  23.4 & 86.8\%       &  26.0 & 96.5\%        &  13.6 & 50.7\% \\
                       &  350                   &   0.3 & 15.8\%       &   1.0 & 60.8\%        &   0.3 & 17.3\% \\[0.2em]
\multirow{1}{0em}{500} &  0                     &  12.0 & 90.5\%       &  13.2 & 99.3\%        &  13.2 & 99.3\% \\
                       &  100                   &  11.6 & 89.2\%       &  12.9 & 99.2\%        &   6.6 & 50.6\% \\
                       &  300                   &   7.8 & 73.2\%       &  10.3 & 97.0\%        &   5.3 & 50.0\% \\
                       &  450                   &   0.1 & 8.2\%        &   0.5 & 61.6\%        &   0.1 & 15.7\% \\
\end{tabular}
\caption{\label{TAB:network} Number of signal events $s$ and corresponding selection efficiencies $\varepsilon$ for the low-mass/high-mass network analysis, the BDT selection of Ref.~\cite{Cornell:2021gut}, and a conventional cut-and-count analysis incorporating static jet veto cuts. Results are provided for the set of mass spectra used for NN training. Recall that the low-mass NN is used for $m_{\tilde{\mu}_{_R}} \leq 400$~GeV, and that the high-mass NN is used otherwise.} 
\end{table}

After NN training, the saved low-mass and high-mass networks are used to analyse the signals emerging from the smuon-neutralino benchmark spectra used in the training phase. This consists of an initial small-scale analysis to determine the expected number of signal and background events passing all cuts, that we denote as $s$ and $b$ respectively. The sensitivity of the NN models can then be inferred from these numbers. We however leave this task to section~\ref{SEC:PROJECTIONS} where a more detailed scan of the parameter space is conducted. Table~\ref{TAB:network} lists the expected number of signal events passing the low-mass and high-mass network selections, respectively, that we compare to predictions obtained using a BDT architecture\footnote{We refer to Ref.~\cite{Cornell:2021gut} for a detailed description of the BDT implementation used as a reference.}, and a conventional cut-and-count analysis with a static jet veto with a threshold of 25~GeV and selections on muons of $p^{\mu_1}_T > 50$~GeV and $p^{\mu_2}_T > 20$~GeV. In addition, we provide the total selection efficiencies. The corresponding number of background events passing the cuts are $b=5.9$ ($\varepsilon=0.8\%$), $2.4$ ($\varepsilon=0.3\%$), $28.4$ ($\varepsilon=4.0\%$) and $33.8$ ($\varepsilon=4.5\%$) for the low-mass network, high-mass network, BDT and static jet veto selection, respectively.

\begin{table}[t!]\centering\setlength\tabcolsep{6pt}\renewcommand{\arraystretch}{1.3}
  \begin{tabular}{P{0.1\linewidth}P{0.1\linewidth}|P{0.1\linewidth}P{0.1\linewidth}P{0.1\linewidth}P{0.1\linewidth}P{0.1\linewidth}P{0.1\linewidth}}
$m_{\tilde{\mu}_{_R}}$ (GeV) & $m_{\tilde{\chi}^0_1}$ (GeV) & $p^{j_1}_T < p^{\mu_1}_T$ & $H_T < p^{\mu_1}_T$ & $p^{j_1}_T < S_T$ & $H_T < S_T$ & $p^{j_1}_T < p^{\mu_2}_T$ & $H_T < p^{\mu_2}_T$\\[0.5em]
\hline
\multirow{1}{0em}{100} &  0  &  79.7~(88.9\%)  &  42.6~(47.6\%) &  87.8~(98.0\%) & 54.2~(60.4\%) &  78.8~(88.0\%)  &  43.2~(48.2\%)\\
                       &  25 &  17.8~(83.8\%)  &   8.9~(96.4\%)  &  20.4~(42.4\%) & 10.7~(50.5\%) &  16.7~(79.3\%) & 8.4~(39.6\%) \\[0.2em]
\multirow{1}{0em}{200} &  0  & 223.3~(93.8\%) & 127.0~(53.4\%) & 236.5~(99.4\%) & 195.1~(81.9\%) & 222.9~(93.6\%) & 126.7~(53.2\%)\\
                       &  50 & 200.1~(93.3\%) & 111.2~(51.8\%) & 212.9~(99.2\%) & 172.1~(80.2\%) & 200.5~(93.4\%) & 111.9~(52.2\%) \\
                       &  100& 124.9~(90.8\%) &  64.0~(46.5\%) & 135.2~(98.3\%) &  97.7~(71.1\%) & 124.3~(90.4\%) &  64.2~(46.7\%) \\
                       &  150&   2.4~(49.9\%) &   0.9~(18.5\%) &   3.6~(76.6\%) &   1.2~(25.9\%) &   2.5~(53.4\%) &   0.9~(19.1\%) \\[0.2em]
\multirow{1}{0em}{300} &  0  &  80.9~(95.4\%) &  53.5~(63.1\%) &  84.5~(99.6\%) &  77.2~(91.0\%) &  80.0~(95.5\%) &  53.4~(63.0\%) \\
                       &  50 &  77.5~(95.3\%) &  50.3~(61.8\%) &  81.0~(99.6\%)  &  73.5~(90.4\%) &  77.2~(95.0\%) &  50.4~(62.0\%) \\
                       &  100&  72.9~(94.5\%) &  45.5~(59.0\%) &  76.7~(99.5\%) &  68.2~(88.4\%) &  72.9~(94.5\%) &  45.6~(59.1\%) \\
                       &  200&  36.6~(89.9\%) &  18.5~(45.5\%) &  40.0~(98.3\%) &  29.4~(72.4\%) &  36.5~(89.7\%) &  18.4~(45.4\%) \\
                       &  250&   1.9~(55.4\%) &   0.7~(20.1\%) &   2.7~(80.1\%) &   1.0~(28.3\%) &   1.9~(56.0\%) &   0.7~(19.9\%) \\[0.2em]
\multirow{1}{0em}{400} &  0  &  30.9~(96.4\%) &  22.7~(70.7\%) &  32.0~(99.8\%) &  30.3~(94.6\%) &  31.0~(96.5\%) &  22.5~(70.3\%) \\
                       &  200&  25.5~(94.7\%) &  16.5~(61.3\%) &  26.8~(99.6\%) &  24.2~(89.8\%) &  25.5~(94.7\%) &  16.4~(61.0\%) \\
                       &  350&   1.0~(57.8\%) &   0.4~(22.0\%) &   1.4~(81.2\%) &   0.5~(30.0\%) &   1.0~(58.2\%) &   0.4~(21.1\%) \\[0.2em]
\multirow{1}{0em}{500} &  0  &  12.9~(97.4\%) &  10.2~(76.5\%) &  13.3~(99.9\%) &  12.8~(96.5\%) &  12.9~(97.3\%) &  10.1~(76.0\%)\\
                       &  100&  12.6~(97.0\%) &   9.7~(74.9\%) &  13.0~(99.8\%) &  12.5~(96.0\%) &  12.6~(96.9\%) &   9.7~(74.7\%) \\
                       &  300&  10.1~(94.5\%) &   6.6~(62.1\%) &  10.6~(99.5\%) &   9.6~(90.4\%) &  10.1~(94.6\%) &   6.7~(62.4\%) \\
                       &  450&   0.5~(58.3\%) &   0.2~(21.4\%) &   0.6~(81.0\%) &   0.5~(30.7\%) &   0.2~(57.5\%) &   0.2~(21.7\%) \\
\end{tabular}
\caption{\label{TAB:DJV} Number of signal events and corresponding signal selection efficiencies, $s(\varepsilon)$, for the six dynamic jet vetoes of Eqn.~\eqref{eq:DJV}. Results are provided for the set of mass spectra used for the training of the low-mass and hig-mass NNs.} 
\end{table}

Ref.~\cite{Fuks:2019iaj} explored the potential of dynamic jet vetoes in light of the smuon signal being typically very complicated to observe. In order to assess the pertinence of these findings in comparison with NN or BDT selections, we present in Tables~\ref{TAB:DJV} the expected number of signal events after several dynamic jet veto cuts. Recall that in contrast to the case of the static jet veto cut, events are rejected if the proxy observables used to assess the leptonic activity has a larger value than the one used to estimate the hadronic activity, on an event-by-event basis. We follow the same strategy as the one proposed in Ref.~\cite{Fuks:2019iaj}, which corresponds to six possibilities for the implementation of a dynamic jet veto. Events are respectively vetoed if they satisfy:
\begin{equation}\label{eq:DJV}
    (a)~p^{j_1}_T > p^{\ell_1}_T\,, \quad
    (b)~H_T > p^{\ell_1}_T\,,\quad
    (c)~p^{j_1}_T > S_T\,,\quad
    (d)~H_T > S_T\,,\quad
    (e)~p^{j_1}_T > p^{\ell_2}_T\,,\quad
    (f)~H_T > p^{\ell_2}_T.
\end{equation}
The expected number of background events passing the dynamic jet veto cuts are $b=204.2$ ($\varepsilon=27.5\%$), $57.4$ ($\varepsilon=7.7\%$), $503.7$ ($\varepsilon=67.8\%$), $125.9$ ($\varepsilon=16.9\%$), $230.8$ ($\varepsilon=31.1\%$) and $125.9$ ($\varepsilon=16.9\%$) for the six veto implementations $(a)$, $(b)$, $(c)$, $(d)$, $(e)$ and $(f)$ of Eqn.~\eqref{eq:DJV}, respectively. 

The above results indicate that we generally get a higher number of expected signal events when using dynamic jet veto cuts than with static jet veto cuts, which is expected given the greater rigidity of the latter threshold (see Ref.~\cite{Fuks:2019iaj} for a more thorough comparison between static and dynamic jet vetoes). Now, comparing two machine learning algorithms, table~\ref{TAB:network} shows that BDTs generally rank higher than NNs and the static jet veto in terms of the fraction of expected signal events that pass the cuts. However, that is at the cost of a less effective background rejection, the hallmark of a model that is optimised for recall. By contrast, our low-mass and high-mass networks are optimised for precision and, as such, are more conservative when identifying signal events. They have the lowest false positives (\textit{i.e.}\ less noise impedes analysis) and, as such, they generally rank highest in sensitivity. We will quantify this in the section~\ref{SEC:PROJECTIONS}, where a detailed scan of the parameter space is achieved.  

\section{Sensitivity and projections for smuon searches at the LHC}
\label{SEC:PROJECTIONS}

Having trained our two networks on the combined signal and background dataset, we can then use them to perform a detailed scan of the parameter space and get an estimate for the associated sensitivity of the LHC. However, the workflow involving {\tt pytorch} networks processes lists of signal data gleaned from many samples, which is very inefficient as we want to test large numbers of signal parameter points. Therefore we exported the two trained models to the {\sc ONNX} format so that they can be used as a final selection cut in a new analysis written for {\sc HackAnalysis}, using the features described in Ref.~\cite{Goodsell:2024aig}. If the network yields a score above the threshold value (0.7 here) then the cut is passed, otherwise it fails. Our analysis is based on that used above for scraping the data, except that instead of writing the information to disk, it tests each specific event and decides whether it populates some of four signal regions with dedicated purposes. The first two regions respectively make use of the high-mass network (``High Network'') and low-mass network (``Low Network'') that we have trained, while for comparison purposes, the last two regions refer to the BDT selection of Ref.~\cite{Cornell:2021gut} (``BDT'') and a cut-and-count selection relying on a static jet veto (``Jet Veto''). 

\begin{figure}
\includegraphics[width=0.6\textwidth]{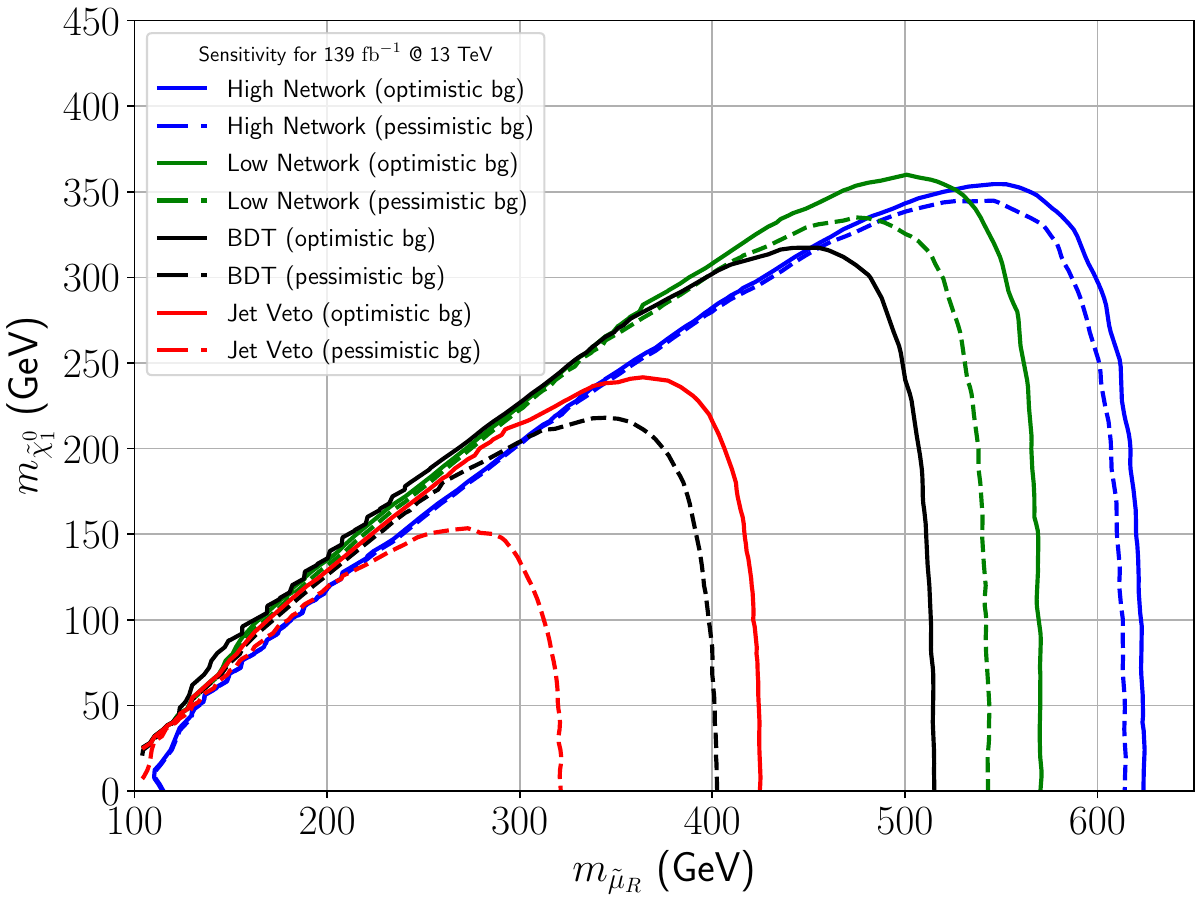}
\caption{\label{FIG:scanlimit}Projected sensitivity of the LHC Run 2 to a smuon signal using our low-mass (green) and high-mass (blue) NN-based analysis, the BDT analysis of Ref.~\cite{Cornell:2021gut} (black) and a conventional cut-and-count an analysis with static jet vetoes (red). We consider an optimistic background hypothesis without any systematic uncertainties, and a pessimistic one where we assume 50\% uncertainty on the background.}
\end{figure}

We use this code to perform an analysis corresponding to the configuration of the LHC Run 2, relying on a luminosity of $139\ \mathrm{fb}^{-1}$ of 13 TeV proton collisions. The sensitivity is provided in the form of expected exclusion contours in figure~\ref{FIG:scanlimit}. For each of the four signal regions, we give exclusion curves for two different background hypotheses, namely an ``optimistic'' one where we neglect any uncertainty in the expected number of background events, and a ``pessimistic'' one where the uncertainty on the background is taken to be $50\%$. Without a detailed investigation of uncertainties, which would be fruitless without an accurate detector simulation and more precision in data simulation, it is not possible to be precise about this figure. However, based on other analyses we may assume that the true background uncertainty should sit well below our pessimistic expectations. 

We can draw several conclusions from figure \ref{FIG:scanlimit}. The first is that separating the signals into two training sets related to `low' and `high' mass BSM spectra is highly effective in helping the network to discriminate the signal from the background, allowing for a substantial gain at high mass. The second is that while the BDT has a higher selection efficiency for signal events, its (much) poorer background rejection power limits its capabilities, especially for high mass signals where the number of events populating the signal regions is smaller by virtue of the smaller corresponding cross sections. Here, the signal selection efficiency is indeed high for both the BDT and the NN, so that  the better background rejection power of the NN yields a gain of about $100$ GeV in reach for the smuon mass (and of even 200 GeV compared to the static jet veto case). These estimates can be compared with associated Run 2 analyses that have been performed by the collaborations. In the initial CMS search for smuons using $35.9\ \mathrm{fb}^{-1}$ and a static jet veto, the limit on the smuon mass only extended to $200$~GeV in the  case of a massless neutralino, as can be seen in the lower-right plot of figure~7 in Ref.~\cite{CMS:2018eqb}. More recent searches using the full Run 2 dataset have been performed, and notably ATLAS found mass limits which very closely resemble our findings for a static jet veto. This can be seen in figure~8(b) from Ref.~\cite{ATLAS:2019lff}, where the mass limits on right-handed smuons extend to $450$~GeV for massless neutralinos. This is again very similar to our estimate for an optimistic background uncertainty. Clearly, then, we could obtain more information from the data and better limits using a machine-learning approach relying on BDTs or NNs. 

\begin{figure}
\includegraphics[width=0.5\textwidth]{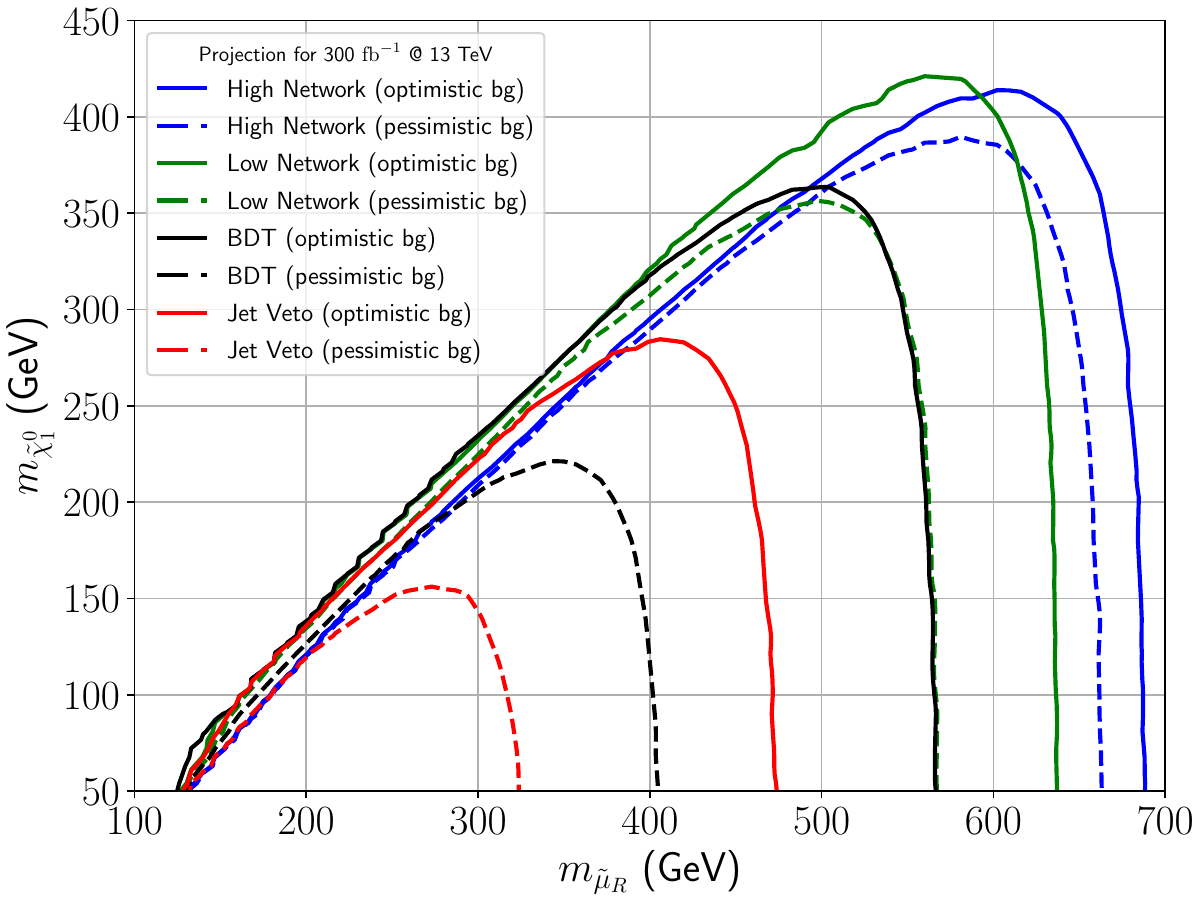}\includegraphics[width=0.5\textwidth]{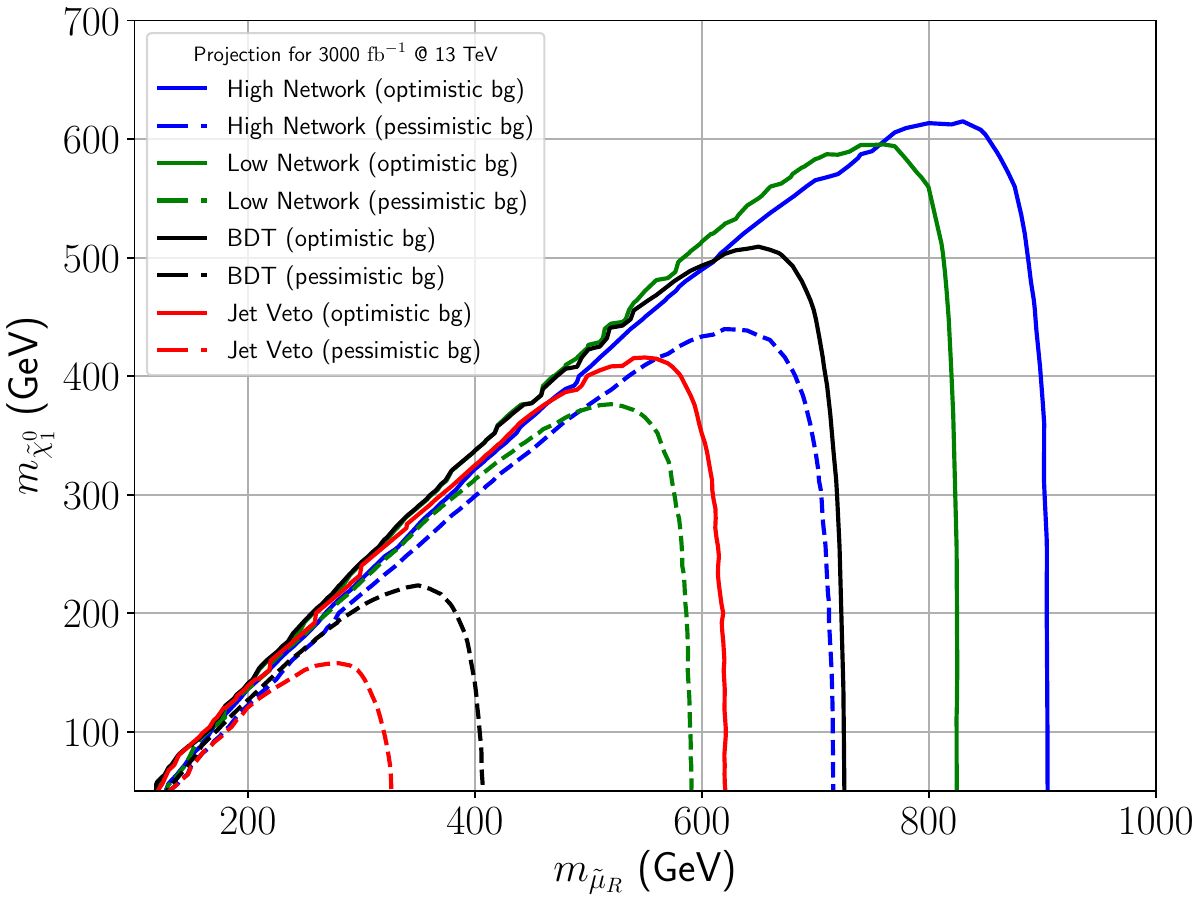}
\caption{\label{FIG:HLlimits}Same as figure~\ref{FIG:scanlimit}, but for a projected luminosity of $300\ \mathrm{fb}^{-1}$ (left) and $3000\ \mathrm{fb}^{-1}$  (right).}
\end{figure}

With an eye to LHC Run 3 and beyond, it is then interesting to consider how the limits change with luminosity. Since the background and training is so resource intensive, we did not re-simulate events at a higher centre-of-mass energy, nor did we retrain our networks for higher-mass samples, but instead considered higher luminosities at $13$ TeV. In figure~\ref{FIG:HLlimits} we present projections for $300\ \mathrm{fb}^{-1}$ (left panel) and $3000\ \mathrm{fb}^{-1}$ (right panel), which correspond to the Run 3 and HL-LHC luminosities respectively. Notably, we see that there is a significant gain in sensitivity for the HL-LHC, with an advantage, in the case of a mass spectrum featuring a massless neutralino, of the NN over the BDT of $200$ GeV, and even $300$ when comparing to the static jet veto analysis. Of course, by retraining the networks with higher-mass samples (the highest-mass sample that we used during the training phase featuring smuons of $500$~GeV) we would expect slightly better performance. It is nevertheless remarkable that the NN shows a good ability to generalise and extrapolate in the large-mass regime. It should however be kept in mind that the potential gain in retraining would be modest, as the signal selection efficiencies are already rather high.

\section{Conclusions}
\label{SEC:CONCLUSIONS}

We have proposed a new LHC search strategy dedicated to finding right-handed smuons that decay to a muon and a neutralino, using a NN discriminator as a final cut. We have shown that using `derived' inputs based on other inputs already provided to the network (such as relative angles between leptons or ratios of leptonic and hadronic observables) gives an improvement in the training speed and performance of the network. Furthermore, we showed that splitting the signal data obtained by combining simulations related to different mass spectra according to the smuon mass can yield different signal selections, with better sensitivities in given mass ranges. This is particularly useful to get better sensitivity to scenarios featuring heavier smuons, associated with rarer signals. We provided estimates for limits that could be obtained at the LHC by using only Run 2 data, together with projections for the LHC Run 3 and HL-LHC luminosities. We have shown that our approach can greatly enhance the reach of the LHC compared to traditional ones relying on conventional cut-and-count analysis employing static jet vetoes. 

In this work, with our ML approach, we are able to approach the fundamental limit of the analysis reach -- namely the point where the cross-section is barely enough to produce a few events -- for the cases where there is a large mass splitting between smuon and neutralino. Yet, the discriminator that we used is relatively simple: underlining the potential of this technology. However, when the neutralino mass approaches that of the smuon, the decay products become softer, and evidently the signal is much harder to distinguish from the background; for that case, and for other models with a more irreducible background, it would be interesting to consider more sophisticated ML architectures. One approach that we initially attempted, and may yet be useful, is to use a mini-batch discrimination layer: this allows information about \emph{batches} of events to be taken into account, which may also help distinguishing signal from background and therefore also speeds up training. However, this would also require a new approach to processing the events. In addition, to significantly gain sensitivity in the degenerate regime, it would be necessary to relax some of the preselection cuts: this would require treating much larger datasets, and would also require a much higher rejection efficiency of the discriminator (it would, at the very least, have to learn to apply the same preselection cuts -- but ideally it would improve upon them).

It is a challenge to use ML techniques to aid LHC analyses because of the need for large amounts of data processing in signal and background generation. We have demonstrated a workflow, for the first time, using publicly available recasting tools that will allow future similar studies to be conducted: large numbers of background and selected signal events can be generated with \hackanalysis, a network can be trained using any ML framework such as {\tt pytorch} that we used, the networks can be exported into the {\tt ONNX} format, and finally fast projections can be obtained for the exclusion plots using a much large number of sample parameter points by treating the analysis as a traditional cut-and-count analysis with the network as the final cut, and using statistics routines from \hackanalysis. To ease adoption of this workflow, we have made all the code for this paper available (with instructions) in a {\tt zenodo.org} record \cite{goodsell_2024_14046530}, including all the training and pseudo-analysis codes. To run the pseudo-analyses, version {\tt 2.2} of \hackanalysis is required, which is available now from its website\footnote{\url{https://goodsell.pages.in2p3.fr/hackanalysis/}}. The same (nearly identical) workflow will soon be possible in \madanalysis once the {\tt ONNX} interface is made publicly available. We hope this development will facilitate similar studies.

\acknowledgments

B.F. and M.D.G. are supported in part by Grant ANR-21-CE31-0013, Project DMwithLLPatLHC, from the \emph{Agence Nationale de la Recherche} (ANR), France. A.S.C. is supported in part by the National Research Foundation (NRF) of South Africa. A.M.N. is supported by an SA-CERN Excellence Bursary through iThemba LABS.

\bibliographystyle{JHEP}
%\bibliography{Bibliography/bibliography}

\bibliography{references}

\end{document}